\documentclass[twocolumn]{aastex631}

\usepackage{devanagari}
\usepackage{upgreek}
\usepackage[encapsulated]{CJK} % for chinese characters
\usepackage{apjfonts}
\usepackage{ucs}
\newcommand{\cntext}[1]{\begin{CJK}{UTF8}{gbsn}#1\end{CJK}}

\providecommand{\sorthelp}[1]{}

\received{April 14, 2023}
\revised{5 August, 2023}
\accepted{7 August, 2023}
\published{29 August, 2023}
\submitjournal{PSJ}

\shorttitle{Microwave Observations of Venus with CLASS}
\shortauthors{Dahal et al.}

\begin{document}

\title{Microwave Observations of Venus with CLASS}

\correspondingauthor{Sumit Dahal}
\email{sumit.dahal@nasa.gov}

\author[0000-0002-1708-5464]{Sumit Dahal}
\affiliation{NASA Goddard Space Flight Center, 8800 Greenbelt Road, Greenbelt, MD 20771, USA}
\affiliation{The William H. Miller III Department of Physics and Astronomy, Johns Hopkins University, 3701 San Martin Drive, Baltimore, MD 21218, USA}

\author{Michael K. Brewer}
\affiliation{The William H. Miller III Department of Physics and Astronomy, Johns Hopkins University, 3701 San Martin Drive, Baltimore, MD 21218, USA}

\author[0000-0001-8379-1909]{Alex B. Akins}
\affiliation{Jet Propulsion Laboratory, California Institute of Technology, Instruments Division, Pasadena, CA 91011, USA}

\author[0000-0002-8412-630X]{John W. Appel}
\affiliation{The William H. Miller III Department of Physics and Astronomy, Johns Hopkins University, 3701 San Martin Drive, Baltimore, MD 21218, USA}

\author[0000-0001-8839-7206]{Charles L. Bennett}
\affiliation{The William H. Miller III Department of Physics and Astronomy, Johns Hopkins University, 3701 San Martin Drive, Baltimore, MD 21218, USA}

\author[0000-0001-8468-9391]{Ricardo Bustos}
\affiliation{Facultad de Ingenier\'{i}a, Universidad Cat\'{o}lica de la Sant\'{i}sima Concepci\'{o}n, Alonso de Ribera 2850, Concepci\'{o}n, Chile}

\author[0000-0002-7271-0525]{Joseph Cleary}
\affiliation{The William H. Miller III Department of Physics and Astronomy, Johns Hopkins University, 3701 San Martin Drive, Baltimore, MD 21218, USA}

\author[0000-0002-0552-3754]{Jullianna D. Couto}
\affiliation{The William H. Miller III Department of Physics and Astronomy, Johns Hopkins University, 3701 San Martin Drive, Baltimore, MD 21218, USA}

\author[0000-0003-3853-8757]{Rahul Datta}
\affiliation{The William H. Miller III Department of Physics and Astronomy, Johns Hopkins University, 3701 San Martin Drive, Baltimore, MD 21218, USA}

\author[0000-0001-6976-180X]{Joseph Eimer}
\affiliation{The William H. Miller III Department of Physics and Astronomy, Johns Hopkins University, 3701 San Martin Drive, Baltimore, MD 21218, USA}

\author[0000-0002-4782-3851]{Thomas Essinger-Hileman}
\affiliation{NASA Goddard Space Flight Center, 8800 Greenbelt Road, Greenbelt, MD 20771, USA}
\affiliation{The William H. Miller III Department of Physics and Astronomy, Johns Hopkins University, 3701 San Martin Drive, Baltimore, MD 21218, USA}

\author[0000-0001-7466-0317]{Jeffrey Iuliano}
\affiliation{Department of Physics and Astronomy, University of Pennsylvania, 209 South 33rd Street, Philadelphia, PA 19104, USA}
\affiliation{The William H. Miller III Department of Physics and Astronomy, Johns Hopkins University, 3701 San Martin Drive, Baltimore, MD 21218, USA}

\author[0000-0002-4820-1122]{Yunyang Li (\cntext{李云炀}$\!\!$)}
\affiliation{The William H. Miller III Department of Physics and Astronomy, Johns Hopkins University, 3701 San Martin Drive, Baltimore, MD 21218, USA}

\author[0000-0003-4496-6520]{Tobias~A. Marriage}
\affiliation{The William H. Miller III Department of Physics and Astronomy, Johns Hopkins University, 3701 San Martin Drive, Baltimore, MD 21218, USA}

\author[0000-0002-5247-2523]{Carolina N\'{u}\~{n}ez}
\affiliation{The William H. Miller III Department of Physics and Astronomy, Johns Hopkins University, 3701 San Martin Drive, Baltimore, MD 21218, USA}

\author[0000-0002-4436-4215]{Matthew~A. Petroff}
\affiliation{Center for Astrophysics, Harvard \& Smithsonian, 60 Garden Street, Cambridge, MA 02138, USA}
\affiliation{The William H. Miller III Department of Physics and Astronomy, Johns Hopkins University, 3701 San Martin Drive, Baltimore, MD 21218, USA}

\author[0000-0001-5704-271X]{Rodrigo Reeves}
\affiliation{CePIA, Departamento de Astronom\'{i}a, Universidad de Concepci\'{o}n, Concepci\'{o}n, Chile}

\author[0000-0003-4189-0700]{Karwan Rostem}
\affiliation{NASA Goddard Space Flight Center, 8800 Greenbelt Road, Greenbelt, MD 20771, USA}

\author[0000-0001-7458-6946]{Rui Shi (\cntext{时瑞}$\!\!$)}
\affiliation{The William H. Miller III Department of Physics and Astronomy, Johns Hopkins University, 3701 San Martin Drive, Baltimore, MD 21218, USA}

\author[0000-0003-3487-2811]{Deniz A. N. Valle}
\affiliation{The William H. Miller III Department of Physics and Astronomy, Johns Hopkins University, 3701 San Martin Drive, Baltimore, MD 21218, USA}

\author[0000-0002-5437-6121]{Duncan J. Watts}
\affiliation{Institute of Theoretical Astrophysics, University of Oslo, P.O. Box 1029 Blindern, N-0315 Oslo, Norway}
\affiliation{The William H. Miller III Department of Physics and Astronomy, Johns Hopkins University, 3701 San Martin Drive, Baltimore, MD 21218, USA}

\author[0000-0003-3017-3474]{Janet L. Weiland}
\affiliation{The William H. Miller III Department of Physics and Astronomy, Johns Hopkins University, 3701 San Martin Drive, Baltimore, MD 21218, USA}

\author[0000-0002-7567-4451]{Edward J. Wollack}
\affiliation{NASA Goddard Space Flight Center, 8800 Greenbelt Road, Greenbelt, MD 20771, USA}

\author[0000-0001-5112-2567]{Zhilei Xu (\cntext{徐智磊}$\!\!$)}
\affiliation{MIT Kavli Institute, Massachusetts Institute of Technology, 77 Massachusetts Avenue, Cambridge, MA 02139, USA}
\affiliation{The William H. Miller III Department of Physics and Astronomy, Johns Hopkins University, 3701 San Martin Drive, Baltimore, MD 21218, USA}

\begin{abstract}
We report on the disk-averaged absolute brightness temperatures of Venus measured at four microwave frequency bands with the Cosmology Large Angular Scale Surveyor (CLASS). We measure temperatures of \mbox{432.3 $\pm$ 2.8 K}, \mbox{355.6 $\pm$ 1.3 K}, \mbox{317.9 $\pm$ 1.7 K}, and \mbox{294.7 $\pm$ 1.9 K} for frequency bands centered at 38.8, 93.7, 147.9, and 217.5~GHz, respectively. We do not observe any dependence of the measured brightness temperatures on solar illumination for all four frequency bands. A joint analysis of our measurements with lower frequency Very Large Array (VLA) observations suggests relatively warmer ($\sim$ 7 K higher) mean atmospheric temperatures and lower abundances of microwave continuum absorbers than those inferred from prior radio occultation measurements.

\end{abstract}

\keywords{\href{http://astrothesaurus.org/uat/1763}{Venus (1763)}; \href{http://astrothesaurus.org/uat/182}{Brightness temperature (182)}; \href{http://astrothesaurus.org/uat/2120}{Atmospheric composition (2120)}}

\section{Introduction}\label{sec:intro}
Since the late 1950s, several spacecraft and earth-based observatories have probed the Venusian surface and atmosphere at various radio wavelengths \citep{mayer1958, barrett1964, pollack1967, dePater1990, pettengill1992, butler2001}. These observations have established that Venus has a hot surface ($\sim$ 750 K) surrounded by a very thick atmosphere, primarily consisting of CO$_2$ ($\sim$ 96\%) with a small amount of N$_2$ and trace amounts of other molecules like SO$_2$ and H$_2$SO$_4$ \citep{muhleman1979,oyama1979}. While radio wavelengths longer than a few cm probe the hot Venusian surface, decreasing wavelengths successively probe increasing altitudes in the atmosphere \citep{butler2001, akinsthesis}. The atmospheric gases and aerosols provide significant microwave opacity, resulting in a steep temperature decrease in the spectrum at shorter wavelengths. An accurate measurement of the Venus microwave brightness temperature spectrum can therefore provide valuable information about the composition
and dynamics of various layers of its atmosphere. In this paper, we report on the disk-averaged brightness temperatures of Venus at four microwave bands, measured with the Cosmology Large Angular Scale Surveyor (CLASS; \citealt{tom2014,katie2016}).

CLASS is an array of microwave polarimeters that surveys 75\% of the sky every day from the Atacama Desert at four frequency bands centered near 40, 90, 150, and 220 GHz. All CLASS telescopes use feedhorn-coupled transition-edge sensor (TES) bolometers cooled to temperatures $\lesssim 60$ mK to make high-sensitivity measurements \citep{dahal2022} of microwave sources on the sky. This paper is a follow up to \citet{dahal2021}, where 
the most precise Venus brightness temperature measurements to date in the \textit{Q} and \textit{W} frequency bands centered near 40 and 90 GHz, respectively were presented. Since then, a dichroic \textit{G}-band (150/220 GHz) instrument \citep{dahal2020} has been added to CLASS. We describe the Venus observations performed with the CLASS G-band instrument in Section \ref{sec:class}. In Section~\ref{sec:results}, we present the results from our brightness temperature measurements and examine the phase dependence of the measured temperatures. In Section \ref{sec:model}, we discuss Venus atmospheric modeling. Section \ref{sec:discussion} presents an empirically-perturbed model that is consistent with our observations. Finally, we provide a summary in Section~\ref{sec:summary}.

\section{CLASS Observations} \label{sec:class}
CLASS is designed to make precise measurements of the cosmic microwave background (CMB) polarization on large angular scales. During the nominal survey mode, CLASS scans the microwave sky azimuthally at 45$^\circ$ elevation from a site located at $22^\circ 58^\prime$S latitude and $67^\circ 47^\prime$W longitude with an altitude of approximately 5200 m. Periodically, CLASS performs dedicated observations of bright sources -- the Moon, Venus, and Jupiter -- to calibrate the detector response, obtain telescope pointing information, and characterize the instrument beam \citep{datta2022,xu2020}. During the dedicated Moon/planet observations, the telescopes scan across the source in azimuth at a fixed elevation as the source rises or sets through the telescopes' fields of view. In \citet{dahal2021}, we used the dedicated planet observations to obtain the brightness temperatures of Venus at \textit{Q} and  \textit{W} bands, using Jupiter as a calibration source. Following the same procedure, we extend our brightness temperature measurements to two higher CLASS frequency bands centered near 150 and 220~GHz in this paper.

Between 2021 November 13 and 2022 February 24, the CLASS dichroic \textit{G}-band telescope performed 65 dedicated Venus scans. The same \textit{G}-band instrument configuration observed Jupiter 59 times between 2022 September 26 and 2022 November 7. For each of these observations, we combine the acquired time-ordered data (TOD) with the telescope pointing information to generate planet-centered maps for each detector on the focal plane. The raw detector TOD is calibrated to the measured optical power through detector current versus voltage ($I$-$V$) measurements acquired prior to each observation. We use a robust binned $I$-$V$ calibration method described in \citet{appel2022} with a 1\% median error in per-detector calibration across all observations. 

For each of the planet-centered maps, the measured optical power is corrected for atmospheric transmission to account for the effect of precipitable water vapor (PWV) at the CLASS site \citep{pardo2001}. We use the detector optical loading obtained from $I$-$V$ measurements to estimate the PWV at the CLASS site. The relationship between the PWV and the detector optical loading is described in \citet{appel2022}. We verify that the derived brightness temperatures (Section~\ref{sec:phase}) show no dependence on PWV within the measurement uncertainties, increasing our confidence in the atmospheric opacity correction.

Since the angular diameters ($\lesssim$ 1$^\prime$) for both Venus and Jupiter are much smaller than the telescope beam sizes (FWHM of 23$^\prime$ for 150~GHz and 16$^\prime$ for 220~GHz), the planets can be approximated as point sources for CLASS telescopes. Therefore, following \citet{page2003a}, the brightness temperature of the planet $T_\mathrm{p}$ can be calculated as:
\begin{equation}
   T_\mathrm{p} = T_\mathrm{m} \times (\Omega_\mathrm{B} /\Omega_\mathrm{p}),
\label{eq:temp}
\end{equation}
where $\Omega_\mathrm{p}$ is the solid angle subtended by the planet, $\Omega_\mathrm{B}$ is the telescope beam solid angle, and $T_\mathrm{m}$ is the measured peak detector response (amplitude of an elliptical Gaussian fit to the data) from the planet-centered maps after correcting for the atmospheric transmission during the observations. Refer to \citet{datta2022} and \citet{xu2020} for further details on data acquisition and map-making used to obtain $T_\mathrm{m}$ from dedicated CLASS observations.

\section{Results}\label{sec:results}
During the observing campaign, 390 detectors for 150~GHz and 209 detectors for 220~GHz detected both planets at least 20 times. We analyze these observations in two different ways: (1) per-detector averaging of the respective planet observations to constrain the Venus brightness temperature (Section~\ref{sec:temperature}), and (2) per-observation averaging of the respective detector arrays to examine the phase dependence of the measured temperatures (Section~\ref{sec:phase}).

\subsection{Brightness Temperature}\label{sec:temperature}
\begin{figure*}[ht]
\begin{center}
\includegraphics[scale=0.465]{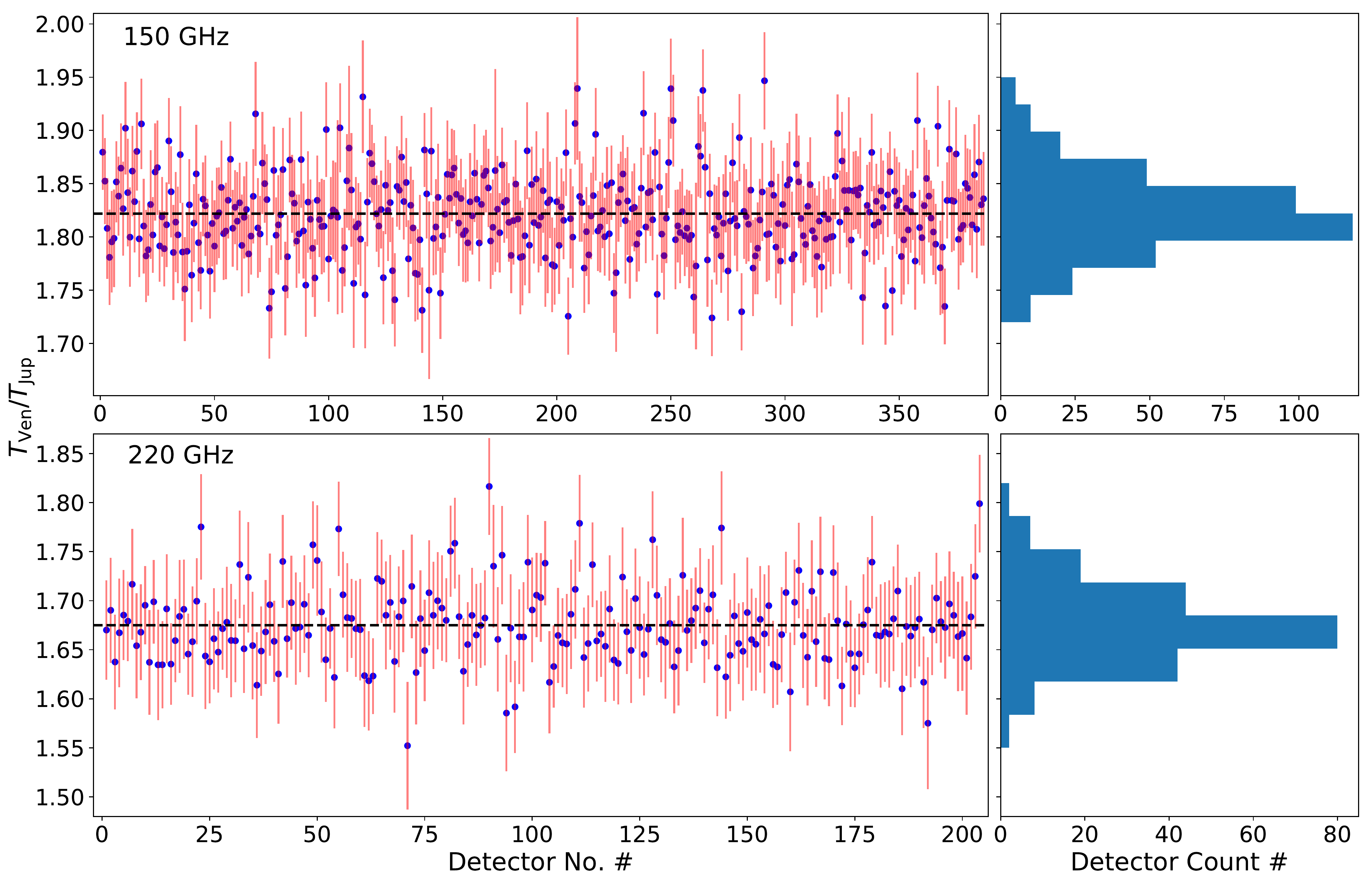}
\end{center}
\caption{The $T_\mathrm{Ven}$/$T_\mathrm{Jup}$ ratio measurements (left) and the corresponding histograms (right) from the CLASS dichroic \textit{G}-band instrument. The uncertainties (red bars) in the measured ratios (blue dots) are the combined errors obtained from the variance of baseline measurements away from the planets. The inverse-variance weighted mean ratios (dashed lines) are 1.822 $\pm$ 0.002 and 1.675 $\pm$ 0.003 for 150 and 220 GHz, respectively, where the uncertainties are the standard errors on the mean.}
\label{fig:venus_temp}
\end{figure*}
For every operating detector on the \textit{G}-band focal plane, we obtain an aggregate planet-centered map by averaging individual maps relative to a fiducial solid angle $\Omega_\mathrm{ref}$. This is performed by scaling the PWV-corrected detector response $T_\mathrm{m}$ by a factor of $\Omega_\mathrm{ref}/\Omega_\mathrm{p}$ while averaging. Since the choice of $\Omega_\mathrm{ref}$ does not affect our final results (see Equation \ref{eq:temp_ratio}), we arbitrarily set  $\Omega_\mathrm{ref}$ = 3.8 $\times$~10$^{-8}$~sr (i.e., 45.45$^{\prime\prime}$ angular diameter) for averaging the maps. For a given observation, $\Omega_\mathrm{p}$ is determined using the distance to the planet with a fixed disk radius $R$. As discussed in \citet{dahal2021}, we use $R$ = 6120 km for Venus and $R$ = 69140 km for Jupiter. The latter is an ``effective R'' for the projected area of Jupiter's oblate disk \citep{weiland2011} calculated using the average Jupiter sub-Earth latitude of $2.81^\circ$ during the observing campaign.

Since both Venus and Jupiter maps are averaged relative to the same $\Omega_\mathrm{ref}$, the ratio of their brightness temperatures (Equation \ref{eq:temp}) for a given detector reduces to:
\begin{equation}
\frac{T_\mathrm{p=Ven}}{T_\mathrm{p=Jup}}= \frac{T_\mathrm{m}^\mathrm{Ven,ref} \times (\Omega_\mathrm{B} /\Omega_\mathrm{ref})}{T_\mathrm{m}^\mathrm{Jup,ref} \times (\Omega_\mathrm{B} /\Omega_\mathrm{ref})} = \frac{T_\mathrm{m}^\mathrm{Ven,ref}}{T_\mathrm{m}^\mathrm{Jup,ref}}.
\label{eq:temp_ratio}    
\end{equation}
Equation \ref{eq:temp_ratio} shows that the brightness temperature ratio is simply the ratio of measured peak responses when scaled to the same $\Omega_\mathrm{ref}$ and does not depend on individual detector properties like $\Omega_\mathrm{B}$. For simplicity, we will refer to the individual planet brightness temperatures $T_\mathrm{p=Ven}$ and $T_\mathrm{p=Jup}$ as $T_\mathrm{Ven}$ and $T_\mathrm{Jup}$, respectively.

Figure \ref{fig:venus_temp} shows the $T_\mathrm{Ven}$/$T_\mathrm{Jup}$ ratios derived from 387 detectors for 150~GHz and 204 detectors for 220~GHz in the CLASS \textit{G}-band instrument. The uncertainties in the ratios are the combined errors obtained from the variance of baseline measurements away from the source for both the Venus and Jupiter maps. For this analysis, we discarded 3 outliers (out of 390) for 150~GHz and 5 (out of 209) for 220~GHz with ratios outside 3 standard deviations from the mean of their respective distributions. For the distributions shown in Figure \ref{fig:venus_temp}, the inverse-variance weighted mean ratios are \mbox{1.822 $\pm$ 0.002} and \mbox{1.675 $\pm$ 0.003} for 150 and 220~GHz, respectively, where the uncertainties are the standard errors on the mean. To verify that these errors represent the uncertainties in the mean of the underlying distribution, we use bootstrapping to generate $10^6$ resamples. For both 150 and 220~GHz, the standard deviation of the mean values of the bootstrapped resamples is the same as the standard error calculated from the parent distribution shown in Figure \ref{fig:venus_temp}.   

\begin{deluxetable}{rrrr}[ht]
\tablecaption{ \label{tab:summary} Summary of CLASS Measurements}
\tablehead{
\colhead{$\nu_\mathrm{e}^\mathrm{RJ}$ [GHz]\tablenotemark{a}} & \colhead{$T_\mathrm{Ven}$/$T_\mathrm{Jup}$} & \colhead{$T_\mathrm{Jup}$ [K]} & \colhead{$T_\mathrm{Ven}$ [K]\tablenotemark{b}}
}
\startdata
38.8 $\pm$ 0.5 & 2.821 $\pm$ 0.015 & 152.6 $\pm$ 0.6\tablenotemark{c} & 430.4 $\pm$ 2.8 \\
93.7 $\pm$ 0.8 & 2.051 $\pm$ 0.004 & 172.8 $\pm$ 0.5\tablenotemark{c} & 354.5 $\pm$ 1.3 \\
147.9 $\pm$ 1.0 & 1.822 $\pm$ 0.002 & 174.2 $\pm$ 0.9\tablenotemark{d} & 317.3 $\pm$ 1.7  \\
217.5 $\pm$ 1.0 & 1.675 $\pm$ 0.003 & 175.8 $\pm$ 1.1\tablenotemark{d} & 294.5 $\pm$ 1.9  \\
\enddata
\tablenotetext{a}{Effective Rayleigh-Jeans point-source center frequencies; see \citet{dahal2022}}
\tablenotetext{b}{Temperature values with respect to blank sky; absolute brightness temperatures can be obtained by adding the RJ temperatures of the CMB of 1.9, 1.1, 0.6, and 0.2 K at 38.8, 93.7, 147.9, and 217.5 GHz, respectively, calculated using the 2.725~K blackbody temperature of the CMB \citep{fixen2009}.
\tablenotetext{c}{Obtained from \citet{bennett2013};  includes 1.7~K correction for 38.8 GHz (see \citealt{dahal2021} for details)}
\tablenotetext{d}{Obtained from \citet{planck2016-LII}; includes 0.1~K correction for 147.9 GHz}
}
\end{deluxetable}

To obtain $T_\mathrm{Ven}$, we multiply the mean CLASS-measured $T_\mathrm{Ven}$/$T_\mathrm{Jup}$ ratios by the corresponding $T_\mathrm{Jup}$ values measured by the Planck satellite \citep{planck2016-LII}. For the higher CLASS \textit{G} band with effective Rayleigh-Jeans (RJ) point-source center frequency of 217.5 $\pm$ 1.0 GHz \citep{dahal2022}, we use $T_\mathrm{Jup}$ = 175.8 $\pm$ 1.1 K from the Planck 217 GHz measurement. For the lower \textit{G} band with effective center frequency of \mbox{147.9 $\pm$ 1.0 GHz}, we use \mbox{$T_\mathrm{Jup}$ = 174.2 $\pm$ 0.9 K}, which is 0.1 K higher than the Planck measurement at 143~GHz. This 0.1~K correction takes into account the difference between the CLASS and Planck center frequencies and is obtained through a local power-law fit between the two Planck $T_\mathrm{Jup}$ values at 143 and 217 GHz. Given a relatively flat $T_\mathrm{Jup}$ spectrum at \textit{G}~band, we obtain the same correction when extrapolating a power-law fit from Planck 100 and 143~GHz measurements as well. The nominal RadioBEAR model\footnote{\url{https://github.com/david-deboer/radiobear}} yields a slightly higher correction of 0.17~K, but we find this to be less reliable as the model spectrum is $\sim 2-3$~K higher than Planck measurements in this frequency range. While both corrections are well within the $T_\mathrm{Jup}$ uncertainty, we adopt the local power-law obtained correction for further analysis as it is consistent across multi-frequency Planck measurements.

Using these $T_\mathrm{Jup}$ values and the mean CLASS-measured $T_\mathrm{Ven}$/$T_\mathrm{Jup}$ ratios shown in Figure \ref{fig:venus_temp}, we obtain $T_\mathrm{Ven}$ of \mbox{317.3 $\pm$ 1.7 K} and \mbox{294.5 $\pm$ 1.9 K} for frequency bands centered at \mbox{147.9 $\pm$ 1.0 GHz} and \mbox{217.5 $\pm$ 1.0 GHz}, respectively. These $T_\mathrm{Ven}$ values represent the disk-averaged Venus brightness temperatures measured with respect to blank sky. The absolute brightness temperatures can be obtained by adding the RJ temperatures of the CMB (0.6~K at 147.9~GHz and 0.2~K at 217.5~GHz), resulting in \mbox{317.9 $\pm$ 1.7 K} and \mbox{294.7 $\pm$ 1.9 K}, respectively. Table \ref{tab:summary} summarizes the CLASS measurements of the Venus brightness temperatures at four microwave frequency bands, including the measurements at two lower bands presented in \citet{dahal2021}. It is worth noting that the $T_\mathrm{Jup}$ values in Table \ref{tab:summary} used to calibrate our $T_\mathrm{Ven}$ measurements are mean disk-integrated brightness temperatures obtained from multi-year WMAP/Planck observations. While temporal temperature variabilities have been reported for different Jovian latitude bands \citep{orton2023}, the disk-averaged temperatures at the frequency bands presented here were found to be stable within the reported uncertainties over nine years of WMAP \citep{bennett2013} and four years of Planck \citep{planck2016-LII} observing seasons. 

\subsection{Phase}\label{sec:phase}
In Section \ref{sec:temperature}, we averaged all the individual planet observations, increasing the signal-to-noise ratio of per-detector aggregate maps to better constrain the Venus brightness temperature. Here, we calculate an array-averaged brightness temperature value for every dedicated Venus observation to examine the phase dependence of the measured temperatures. For a given detector, the denominator value in Equation \ref{eq:temp_ratio} remains the same, but now we calculate the $T_\mathrm{Ven}$/$T_\mathrm{Jup}$ ratio and thus the $T_\mathrm{Ven}$ value separately for each Venus observation. Finally, we average the $T_\mathrm{Ven}$ values obtained from all the detectors in the array for each observation.

\begin{figure}[ht]
\begin{center}
\includegraphics[scale=0.465]{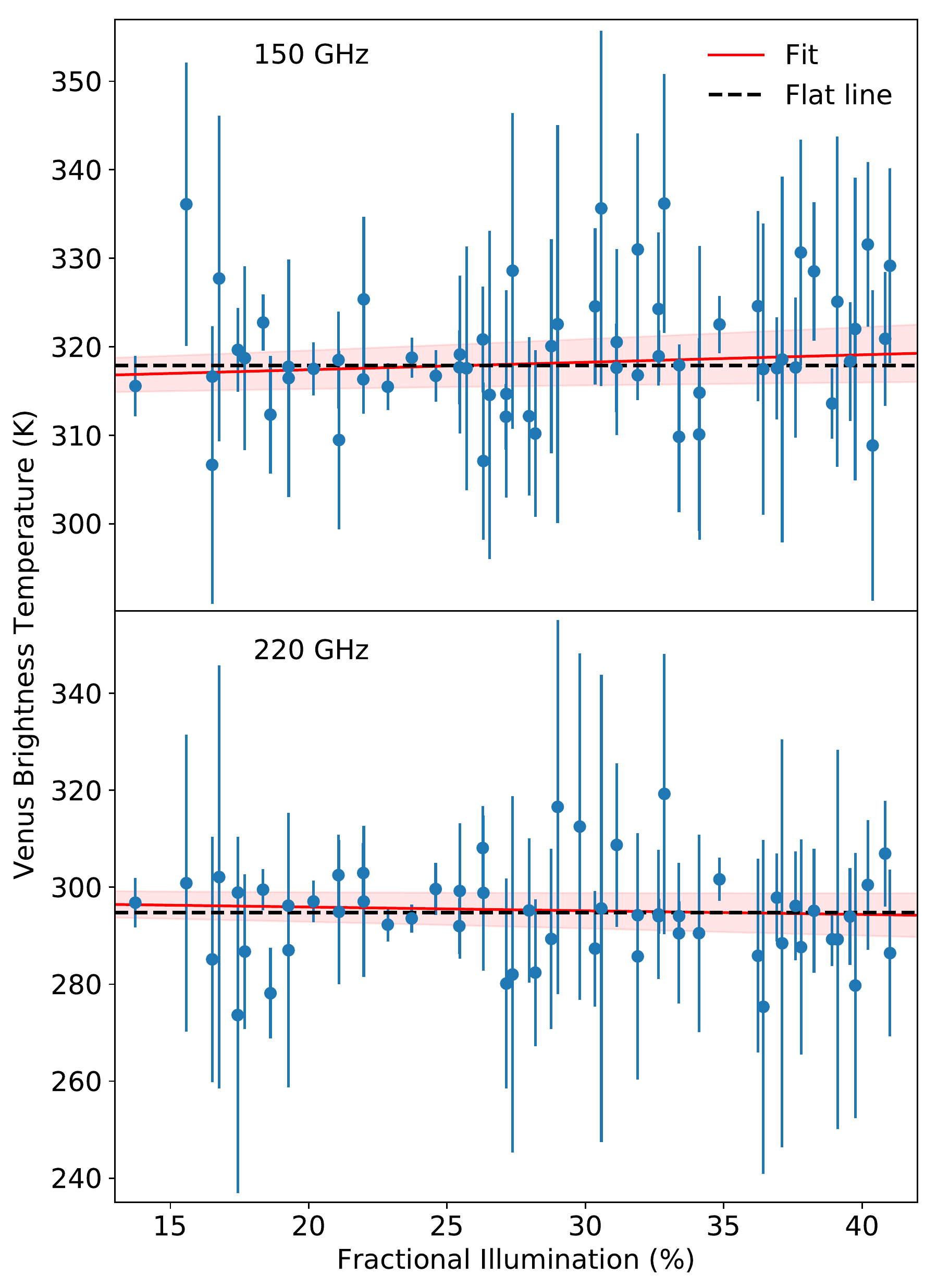}
\end{center}
\caption{The measured array-averaged Venus brightness temperature versus fractional solar illumination. The uncertainties on the data points are the standard errors on the mean. The flat lines (dashed) centered at the absolute brightness temperature values from Table \ref{tab:summary} fall within the 1$\sigma$ uncertainty (shaded red) of the linear best fit lines (solid red). There is no statistically-significant phase dependence of the measured temperatures at both frequency bands.}
\label{fig:venus_phase}
\end{figure}

Figure \ref{fig:venus_phase} shows the array-averaged $T_\mathrm{Ven}$ versus fractional solar illumination (phase) of Venus during the observations. During the CLASS \textit{G}-band observing campaign, the solar illumination of Venus varied from 13.7\% to 41.0\%. However, we do not see a statistically-significant phase dependence of the observed brightness temperatures at either frequency band. The linear fit lines with gradients of \mbox{$0.08 \pm 0.07$ K/\%} for 150~GHz and \mbox{$-0.08 \pm 0.09$ K/\%} for 220~GHz are statistically consistent with being flat. This result is consistent with the absence of phase variation observed at the two lower CLASS frequency bands \citep{dahal2021} and the range of phase-dependent temperature variations inferred from prior radio occultation measurements \citep{tellmann2009}.

\section{Atmospheric Modeling}\label{sec:model}
\begin{figure*}[ht]
\begin{center}
\includegraphics[scale=0.6]{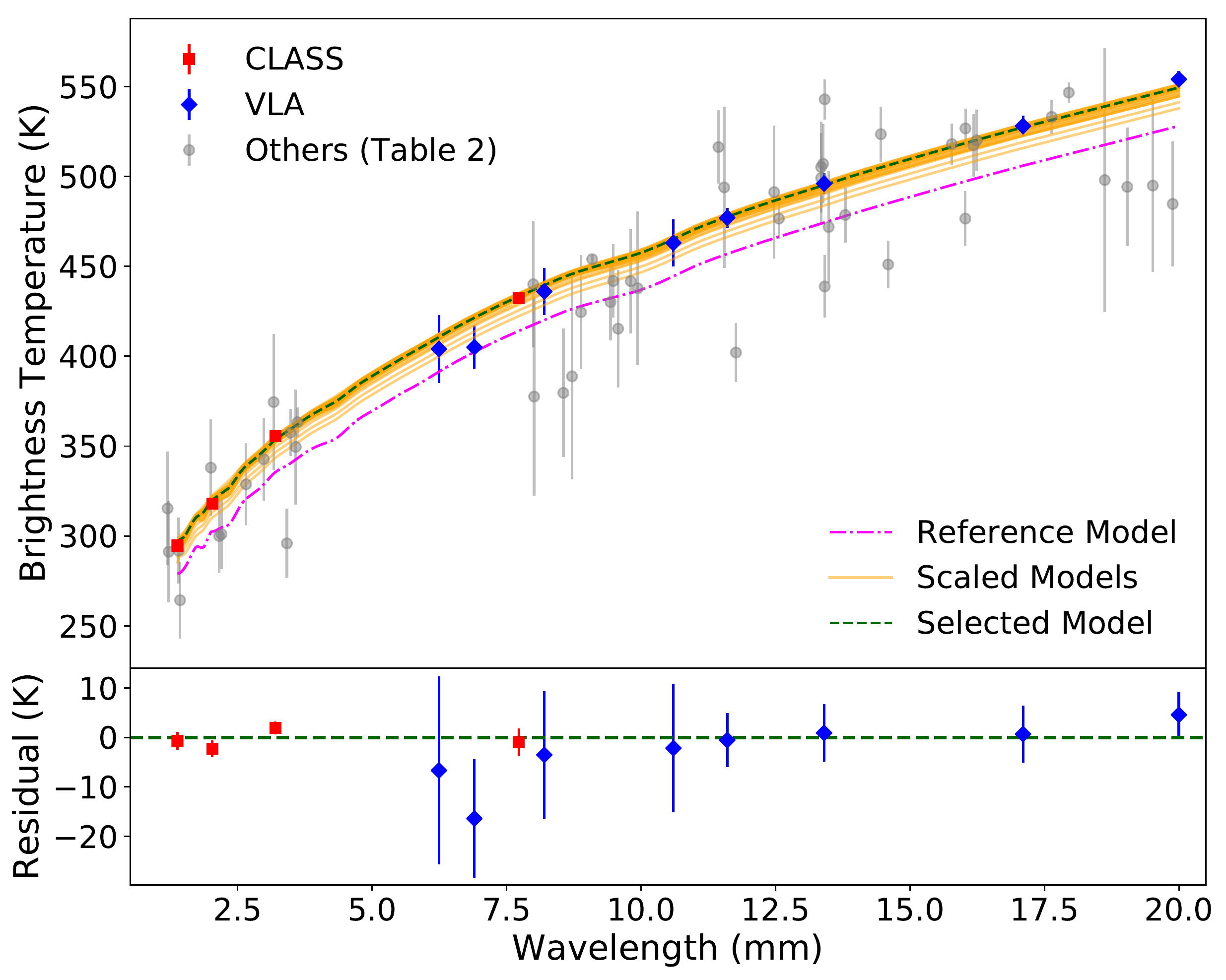}
\end{center}
\caption{Disk-averaged Venus brightness temperature measurements at 1 mm -- 2 cm wavelengths. The data points along with their references are listed in Table \ref{tab:observations}. The magenta dashed-dotted curve is the reference Venusian atmospheric model determined from the radio occultation data, and the solid orange curves are obtained by scaling various model parameters (see Section \ref{sec:discussion} for details) from the reference model. The green dashed curve shows the scaled model that is most consistent with the CLASS and VLA measurements, which are among the most precise measurements to date in this wavelength range.}
\label{fig:venus_spectra}
\end{figure*}

The CLASS observations of the disk-averaged Venus brightness temperatures presented here are the most precise measurements to date at these frequency bands. Figure \ref{fig:venus_spectra} and Table \ref{tab:observations} show the CLASS measurements in context with other published microwave observations. Given that the wavelengths shorter than $\sim$2 cm are primarily sensitive to Venusian atmospheric emission, precise microwave observations can be used to study the composition and dynamics of various layers of the Venusian atmosphere. Here, we perform a joint analysis of the CLASS observations from \mbox{$\sim$1 mm -- 8 mm}  and the Very Large Array (VLA) measurements of \citet{perley2013} from $\sim$7 mm -- 2 cm to obtain constraints on the accuracy of Venus atmospheric composition models.

Despite the precision of the CLASS and VLA brightness temperature measurements, fitting to the disk-averaged spectrum is broadly challenging due to the unresolved nature of the observations and the considerable latitudinal variation in the Venusian atmospheric structure and composition. In principle, it is possible to find an atmospheric model with arbitrary parameters that would produce an exact fit to the measured spectrum. However, the results obtained from such a fit with input parameters that are not physically motivated would not be particularly informative. Therefore, we start with a latitude-dependent atmospheric model informed by radio occultation measurements and other prior analyses, and then scale the model parameters globally to obtain an empirically-perturbed model that is consistent with the CLASS and VLA spectra.

We use a two dimensional, zonally averaged, hemispherically symmetric atmospheric model with 250 m vertical and 5$^\circ$ latitude resolution. The temperature and pressure profiles for the model are taken from the original Venus International Reference Atmosphere (VIRA; \citealt{SEIFF1985}). Between the \mbox{40 -- 90 km} altitudes, the temperature profiles are merged with those determined by \citet{ando2020} from radio occultation refractivity measurements with the Venus Express (VEX; \citealt{Svedhem2007}) and the Akatsuki \citep{Nakamura2016} space probes. The bulk atmospheric composition is 96.5\% CO$_2$ and 3.5\% N$_2$, and trace species including SO$_2$ gas and H$_2$SO$_4$ vapor and aerosol are the primary continuum microwave absorbers. The abundance and spatial distribution of these trace species impact the observed brightness temperature spectrum. For our reference model, we use latitude-dependent SO$_2$ abundance and H$_2$SO$_4$ vapor profiles \citep{OSCHLISNIOK2021} based on the  VEX radio absorption measurements. This SO$_2$ profile has uniform abundance beneath 55 km with rapid depletion above that altitude and features sub-cloud abundances on the order of 50~ppm at lower latitudes, increasing to 150 ppm at the poles. The latitudinally varying H$_2$SO$_4$ vapor profile has maximal abundance values of $\sim$ 12 ppm at equatorial and polar latitudes around 43 -- 47 km altitude. The cloud aerosol mass profiles are taken from the atmospheric transport model of \citet{OSCHLISNIOK2021} which reproduces the VEX H$_2$SO$_4$ vapor distribution well. For the range of aerosol particle sizes inferred from the Pioneer Venus cloud particle size spectrometer (LCPS; \citealt{Knollenberg1980}), we expect the effect from scattering to be negligible as the scattering cross section for these aerosols is multiple orders of magnitude below their absorption cross section at CLASS frequencies \citep{akinsthesis, fahdthesis}. While their impact on the brightness temperature spectrum is expected to be minimal, other species above 1 ppm abundance, specifically H$_2$O, CO, and OCS, are also included at their nominal abundances \citep{KRASNOPOLSKY2007, KRASNOPOLSKY2012}.

The model surface, which primarily affects the brightness temperatures at wavelengths $\gtrsim$ 2 cm, is set to be uniform with a dielectric constant ($\varepsilon_r$) of 4 obtained from the average of the emissivity and reflectivity values determined from the Magellan radar/radiometer observations \citep{pettengill1992}. \citet{JENKINS2002} follow the same approach for their analysis of spatially-resolved VLA observations at wavelengths up to $\sim$ 2 cm, further validating our surface assumption for comparison to the disk-averaged CLASS observations at shorter wavelengths. At these wavelengths, the effect of Bragg and volume scattering by the Venusian surface is negligible. The radiative transfer calculations are equivalent to those described in detail by \citet{butler2001} and \citet{akinsthesis}. The atmosphere is assumed to be locally plane-parallel and optical paths are determined via a ray-tracing approach, which accounts for Venus’ significant atmospheric refraction and incorporates limb-emission effects at the edge of the Venusian disk. The opacities within each homogeneous layer are computed with either continuum or line-by-line opacity models for CO$_2$/N$_2$, SO$_2$, H$_2$SO$_4$ aerosol, and H$_2$SO$_4$ vapor determined from laboratory measurements. We use the models from \citet{fahd1992}, \citet{fahd1991}, and \citet{AKINS2020} for SO$_2$, H$_2$SO$_4$ aerosol, and H$_2$SO$_4$ vapor, respectively. 

Figure \ref{fig:venus_spectra} shows the brightness temperature spectrum obtained from the described model (the magenta dashed–dotted curve). This nominal atmospheric model determined from the radio occultation data produces a spectrum that is colder than the CLASS and VLA measurements, similar to the results from \citet{butler2001} and \citet{JENKINS2002}. To obtain a model that is consistent with the CLASS and VLA measurements, we perturb the nominal reference model 
by uniformly scaling the abundances of the molecular absorbers (multiplicatively) and the temperature profiles (additively).

\begin{deluxetable*}{lll|lll}[ht]
\tablecaption{ \label{tab:observations} Disk-averaged Microwave Venus Brightness Temperature Measurements}
\tablehead{
\colhead{Wavelength [mm]} & \colhead{$T_\mathrm{Ven}$ [K]} &\colhead{Reference} &\colhead{Wavelength [mm]} & \colhead{$T_\mathrm{Ven}$ [K]} &\colhead{Reference}}
\startdata
1.19 & 315.3 $\pm$ 31.6 & \citet{muhleman1979}   & 9.94  & 437.8 $\pm$ 42.9 & \citet{muhleman1979}     \\
1.22 & 291.3 $\pm$ 28.2 & \citet{ulich1974}      & 10.6  & 463 $\pm$ 13     & \citet{perley2013}       \\
\textbf{1.38} & \textbf{294.7 $\pm$ 1.9}  & \textbf{This work}              & 11.44 & 516.3 $\pm$ 20.4 & \citet{muhleman1979}     \\
1.4  & 291.8 $\pm$ 18.4 & \citet{muhleman1979}   & 11.55 & 493.9 $\pm$ 44.9 & \citet{muhleman1979}     \\
1.43 & 264.3 $\pm$ 21.4 & \citet{muhleman1979}   & 11.6  & 477 $\pm$ 5.5    & \citet{perley2013}       \\
2    & 338 $\pm$ 27     & \citet{fasano2021}     & 11.76 & 402 $\pm$ 16.3   & \citet{muhleman1979}     \\
\textbf{2.03} & \textbf{317.9 $\pm$ 1.7}  & \textbf{This work}              & 12.48 & 491.3 $\pm$ 36.9 & \citet{ulich1974}        \\
2.15 & 300 $\pm$ 20.4   & \citet{ulich1974}      & 12.56 & 476.5 $\pm$ 11.2 & \citet{muhleman1979}     \\
2.2  & 301 $\pm$ 19.4   & \citet{muhleman1979}   & 13.35 & 499.1 $\pm$ 25   & \citet{butler2001}       \\
2.65 & 328.7 $\pm$ 23   & \citet{akinsthesis}    & 13.35 & 505.2 $\pm$ 25.3 & \citet{butler2001}       \\
2.98 & 342.7 $\pm$ 23   & \citet{akinsthesis}    & 13.38 & 507 $\pm$ 22     & \citet{steffes1990}      \\
3.17 & 374.5 $\pm$ 37.8 & \citet{muhleman1979}   & 13.4  & 496 $\pm$ 5.8    & \citet{perley2013}       \\
\textbf{3.2}  & \textbf{355.6 $\pm$ 1.3}  & \textbf{\citet{dahal2021}}              & 13.41 & 438.8 $\pm$ 17.3 & \citet{muhleman1979}     \\
3.41 & 295.9 $\pm$ 19.4 & \citet{muhleman1979}   & 13.41 & 542.9 $\pm$ 11.2 & \citet{muhleman1979}     \\
3.48 & 357.5 $\pm$ 13.1 & \citet{ulich1980}      & 13.49 & 471.8 $\pm$ 31.1 & \citet{ulich1974}        \\
3.58 & 349.5 $\pm$ 32   & \citet{ulich1974}      & 13.8  & 478.6 $\pm$ 15.3 & \citet{muhleman1979}     \\
3.61 & 363.3 $\pm$ 8.2  & \citet{muhleman1979}   & 14.46 & 523.5 $\pm$ 15.3 & \citet{muhleman1979}     \\
6.24 & 404 $\pm$ 19     & \citet{perley2013}     & 14.59 & 451 $\pm$ 13.3   & \citet{muhleman1979}     \\
6.9  & 405 $\pm$ 12     & \citet{perley2013}     & 15.78 & 518.0 $\pm$ 11.5 & \citet{rubinomartin2023} \\
\textbf{7.73} & \textbf{432.3 $\pm$ 2.8}  & \textbf{\citet{dahal2021}}              & 16.03 & 476.5 $\pm$ 15.3 & \citet{muhleman1979}     \\
7.99 & 440 $\pm$ 35     & \citet{kolodnerthesis} & 16.03 & 526.7 $\pm$ 10.9 & \citet{rubinomartin2023} \\
8.01 & 377.6 $\pm$ 55.1 & \citet{muhleman1979}   & 16.18 & 517.3 $\pm$ 17.3 & \citet{muhleman1979}     \\
8.2  & 436 $\pm$ 13     & \citet{perley2013}     & 16.24 & 520 $\pm$ 17     & \citet{steffes1990}      \\
8.55 & 379.6 $\pm$ 35.7 & \citet{muhleman1979}   & 17.1  & 528 $\pm$ 5.8    & \citet{perley2013}       \\
8.72 & 388.8 $\pm$ 57.1 & \citet{muhleman1979}   & 17.63 & 533.1 $\pm$ 9.5  & \citet{rubinomartin2023} \\
8.88 & 424.5 $\pm$ 31.6 & \citet{muhleman1979}   & 17.95 & 546.6 $\pm$ 5.7  & \citet{rubinomartin2023} \\
9.08 & 453.6 $\pm$ 3.1  & \citet{hafez2008}      & 18.62 & 498 $\pm$ 73.5   & \citet{muhleman1979}     \\
9.43 & 430.1 $\pm$ 21.4 & \citet{ulich1974}      & 19.04 & 494.2 $\pm$ 33   & \citet{ulich1974}        \\
9.48 & 441.8 $\pm$ 20.4 & \citet{muhleman1979}   & 19.51 & 494.9 $\pm$ 48   & \citet{muhleman1979}     \\
9.57 & 415.3 $\pm$ 32.7 & \citet{muhleman1979}   & 19.88 & 484.7 $\pm$ 34.7 & \citet{muhleman1979}     \\
9.81 & 441.7 $\pm$ 29.1 & \citet{ulich1974}      & 20    & 554 $\pm$ 4.7    & \citet{perley2013}      \\
\enddata
\tablecomments{The bold values highlight the CLASS‐based Venus measurements.}
\end{deluxetable*}

\section{Discussion}\label{sec:discussion}
At shorter millimeter wavelengths, the CLASS measurements are particularly important in constraining the models within the cloud region. For our reference model, however, we find that the shorter-wavelength CLASS measurements are even warmer than those predicted if the only sources of opacity were CO$_2$ and N$_2$. The only way to resolve this discrepancy within the context of the atmospheric model is to increase the magnitude of the physical temperature profile adopted to parameterize the observations in the model. Based on past analyses, it is also not realistic to completely remove SO$_2$ and H$_2$SO$_4$ from the Venusian atmosphere. Therefore, we examine several scaled models with different combinations of increased temperatures and decreased absorber abundances to find a model that is consistent with the CLASS and VLA measurements. For this paper, we explore 28 different models with (1) increases in mean temperature between \mbox{0 -- 8 K}, (2) total SO$_2$,  H$_2$SO$_4$ vapor, and H$_2$SO$_4$ aerosol abundances varied individually in the range between 0.6 -- 1.0 times the nominal values described in Section \ref{sec:model} for the reference model, and (3) an additional cutoff altitude parameter varied between 50 -- 60 km, above which the H$_2$SO$_4$ aerosol and SO$_2$ abundances were set to zero.

The brightness temperature spectra for the 28 scaled models considered in our analysis are shown in Figure \ref{fig:venus_spectra}. The green-dashed curve shows the scaled model with the lowest chi-squared value (reduced chi-square of $\sim$ 1.1 for 7 degrees of freedom). Compared to the reference model, this selected model was obtained by increasing the temperature profile by 7~K, decreasing the SO$_2$, H$_2$SO$_4$ vapor, and H$_2$SO$_4$ aerosol abundances by 30\%, and setting the cutoff altitude to $\sim$ 55~km. While this empirically-perturbed model produces a good fit to the CLASS and VLA brightness temperature measurements, we cannot rule out other models that have slightly warmer temperatures and higher absorber abundances or vice versa. 

Regardless of the particular choice of the best fit model, all of our scaled models that produce a reasonable fit to the CLASS and VLA observations suggest a necessary depletion of microwave opacity within the Venusian middle cloud region. This result is consistent with other observational constraints that SO$_2$ is either chemically depleted within this region or inhibited from diffusive mixing \citep{VANDAELE2017}. Although the exact altitude may vary within a few km, the necessary SO$_2$ depletion altitude of the selected model aligns well with the lower-middle cloud boundary \citep{Knollenberg1980}. A preliminary analysis from \citealt{Noguchi2023} show similar results with lower H$_2$SO$_4$ abundances from Akatsuki radio occultation measurements, compared to those inferred from VEX observations used in our reference model. The temperature increase in our empirically-perturbed model above the clouds is on the order of magnitude expected for diurnal and semi-diurnal variability in cloud-level temperatures \citep{tellmann2009} and near the upper limit for lower atmospheric variability inferred from probe measurements \citep{SEIFF1985}.

\section{Summary}\label{sec:summary}
Using Jupiter as a calibration source, we measure the disk-averaged brightness temperatures of Venus at four microwave frequency bands with CLASS. In \citet{dahal2021}, we had reported brightness temperatures of \mbox{432.3 $\pm$ 2.8 K} and \mbox{355.6 $\pm$ 1.3 K} for frequency bands centered at \mbox{38.8 $\pm$ 0.5 GHz} and \mbox{93.7 $\pm$ 0.8 GHz}, respectively. With the addition of a dichroic \textit{G}-band instrument to CLASS, we measure Venus temperatures of \mbox{317.9 $\pm$ 1.7 K} and \mbox{294.7 $\pm$ 1.9 K} at effective center frequencies of \mbox{147.9 $\pm$ 1.0 GHz} and \mbox{217.5 $\pm$ 1.0 GHz}, respectively. For their respective bands, these CLASS measurements are the most precise disk-averaged Venus brightness temperatures to date. We observe no phase dependence of the measured temperatures at all four frequency bands.

Since the wavelengths below a few cm are sensitive to Venusian atmospheric emission, we perform a joint analysis of the CLASS observations from $\sim$1 mm -- 8~mm  and the VLA measurements from $\sim$7 mm -- 2 cm to obtain constraints on the accuracy of Venus atmospheric composition models. Our analysis suggests the presence of relatively warmer mean atmospheric temperatures (i.e., by $\sim$ 7 K) than those derived from prior radio occultation measurements. In addition, our observations indicate that the abundance of microwave absorbers inferred from the VEX radio occultation measurements could be comparatively overestimated. Further spatially resolved microwave observations of Venus could provide additional context to these disk-integrated observations.

\section*{ACKNOWLEDGMENTS}
\begin{acknowledgments}
We acknowledge the National Science Foundation Division of Astronomical Sciences for their support of CLASS under Grant Numbers 0959349, 1429236, 1636634, 1654494, 2034400, and 2109311. We thank Johns Hopkins University President R. Daniels and the Kreiger School deans for their steadfast support of CLASS. We further acknowledge the very generous support of Jim and Heather Murren (JHU A\&S '88), Matthew Polk (JHU A\&S Physics BS '71), David Nicholson, and Michael Bloomberg (JHU Engineering~'64). CLASS employs detector technology developed in collaboration between JHU and Goddard Space Flight Center under several previous and ongoing NASA grants. Detector development work at JHU was funded by NASA cooperative agreement 80NSSC19M0005. CLASS is located in the Parque Astron\'omico Atacama in northern Chile under the auspices of the Agencia Nacional de Investigaci\'on y Desarrollo (ANID).

We acknowledge scientific and engineering contributions from Max Abitbol, Fletcher Boone, David Carcamo, Ted Grunberg, Saianeesh Haridas, Connor Henley, Lindsay Lowry, Nick Mehrle, Sasha Novack, Bastian Pradenas, Isu Ravi, Gary Rhodes, Daniel Swartz, Bingjie Wang, Qinan Wang, Tiffany Wei, and Zi\'ang Yan. We thank Miguel Angel D\'iaz, Jill Hanson, William Deysher, Joe Zolenas, Mar\'ia Jos\'e Amaral, and Chantal Boisvert for logistical support. We acknowledge productive collaboration with Dean Carpenter and the JHU Physical Sciences Machine Shop team. 

S.D. is supported by an appointment to the NASA Postdoctoral Program at the NASA Goddard Space Flight Center, administered by Oak Ridge Associated Universities under contract with NASA. A. A. acknowledges support from the NASA Solar System Observations program (task order 80NM0018F0612). Contributions by A.A. were carried out at the Jet Propulsion Laboratory, California Institute of Technology, under a contract with NASA (80NM0018D0004). R.R. is supported by the ANID BASAL projects ACE210002 and FB210003. Z. X. is supported by the Gordon and Betty Moore Foundation through grant GBMF5215 to the Massachusetts Institute of Technology.

\end{acknowledgments}

\software{\texttt{PyEphem} \citep{rhodes2011}, \texttt{NumPy} \citep{numpy}, \texttt{SciPy} \citep{scipy}, \texttt{Astropy} \citep{astropy}, \texttt{Matplotlib} \citep{matplotlib}, \texttt{RadioBEAR} \citep{radiobear2,radiobear1}}

\bibliography{venus, Planck_bib}{}
\bibliographystyle{aasjournal}
\end{document}